\documentclass[a4paper,fleqn,usenatbib]{mnras}
\usepackage{amssymb,amsmath,graphicx,psfrag,mathtools}
\usepackage[T1]{fontenc} 
\usepackage{ae,aecompl} 
\usepackage{url}
\title[Do FRB Mark Dark Core Collapse?]{Are Fast Radio Bursts Markers of Dark Core Collapse?}
\author[J. I. Katz]{
J. I. Katz,$^{1}$\thanks{E-mail: katz@wuphys.wustl.edu} 
\\
$^{1}$Department of Physics and McDonnell Center for the Space Sciences,
Washington University, St. Louis, Mo. 63130 USA 
}
\date{Accepted XXX.  Received YYY; in original form ZZZ} 
\pubyear{2017} 
\date{\today}
\begin{document} 
\label{firstpage} 
\pagerange{\pageref{firstpage}--\pageref{lastpage}} 
\maketitle 
\begin{abstract}
Are some neutron stars produced without a supernova, without ejecting mass
in a remnant?  Theoretical calculations of core collapse in massive stars
often predict this.  The observation of the repeating FRB 121102, whose
dispersion measure has not changed over several years, suggests that dark
core collapses are not just failures of computer codes, but may be real.
The existence of one repeating FRB with unchanging dispersion measure is not
conclusive, but within a decade hundreds or thousands of FRB are expected to
be discovered, likely including scores of repeaters, permitting useful
statistical inferences.  A na\"{\i}ve supernova remnant model predicts
observable decline in dispersion measure for 100 years after its formation.
If an upper limit on the decline of 2 pc/cm$^3$-y is set for five repeating
FRB, then the na\"{\i}ve model with nominal parameters is rejected at the
95\% level of confidence.  This may indicate dark neutron star formation
without a supernova or supernova remnant.  This hypothesis may also be
tested with LSST data that would show, if present, a supernova at an
interferometric FRB position if it occurred within the LSST epoch.
\end{abstract}
\section{Introduction}
For several decades, many calculations of core collapse of massive stars
have failed to yield the expected result, an explosion explaining observed
core collapse supernov\ae\ \citep{J17,Mu17,S17}.  Much effort has gone into
improving the calculations to explain the core collapse supernov\ae\ that
surely exist.  Yet this conundrum can be viewed differently:  The na\"{\i}ve
calculations may be predicting a real phenomenon, the formation of neutron
stars without a visible supernova and without expulsion of a envelope
forming a supernova remnant.

We certainly know of core collapse supernov\ae\ that birthed neutron stars,
for their remnants are visible, and some of the supernov\ae\ themselves were
observed in historic times.  Dark neutron star formation is harder to
demonstrate.  If we observe a neutron star, typically a radio pulsar but in
some instances a thermal X-ray source, soft gamma repeater/anomalous X-ray
pulsar or a gamma-ray pulsar, the absence of a surrounding supernova remnant
may be attributed alternatively to dissipation of an older remnant or to
dark neutron star formation.  Demonstrating dark neutron star formation
would require confidence that the neutron star is younger than any plausible
dissipation time of a remnant, one or a few thousand years.  Pulsar spindown
times set upper bounds on neutron star ages, but no pulsar with such a short
spindown time is known to lack a supernova remnant.

The recent discovery \citep{Sp16,Sc16,C17,M17} of the repeating FRB 121102
offers another approach to finding young neutron stars without supernova
remnants.  Over a period of about three years the dispersion measure (DM) of
this FRB has not changed by more than about 5 pc/cm$^3$.  Energetic
arguments \citep{K16} indicate that FRB are associated with young neutron
stars, although it is unclear whether these resemble pulsars, soft gamma
repeaters, or some novel class.  \citet{MKM16,P16,KM17,MBM17} have
considered the constraints that can be placed on the parameters of a
supernova remnant in which FRB 121102 may be embedded.  \citet{PBS17}
suggested that FRB 110220 and 140514, that have consistent positions,
represent a repeating FRB whose DM drastically decreased over three years as
a result of the remnant's expansion.  Most authors assume a very massive
remnant, typically scaling their results to a mass of $10 M_\odot$.

\citet{B17} and \citet{DWY17} suggested that FRB 121102 and the apparently
associated steady radio source \citep{C17} are produced in a pulsar wind
nebula without a confining supernova remnant.  \citet{DWY17} suggested
several possible mechanisms for producing such an object, most of which
involve separating a neutron star from its natal supernova remnant, but also
including accretion-induced collapse of a white dwarf \citep{CS76,NK91}
without expulsion of debris.  This proposed mechanism of dark neutron star
formation may be disparaged by the known association of accreting white
dwarfs with SN Ia.

This note considers the hypothesis that FRB are produced by young neutron
stars formed darkly, without a supernova or a supernova remnant.  Perhaps
some accretion-induced collapses of white dwarfs do not lead to
thermonuclear explosions.  Alternatively, I hypothesize that some core
collapses produce a neutron star without a supernova and without an
expanding remnant.

Within a decade it is likely hundreds or thousands of FRB will be observed, 
including scores of repeaters if they have the same abundance as in the
present database.  This will permit statistical studies of any variation in
dispersion measure; a single repeater with constant dispersion measure may
be a fluke, but if many such are observed the inference will be either that
FRB occur in comparatively old neutron stars whose supernova remnants have
dissipated, or that they are not accompanied by supernova remnants at all.
\section{Why Dark Core Collapse?} 
Two arguments, neither new, lead to the suggestion of dark neutron star
formation.  The first is the difficulty core collapse calculations
\citep{J17,Mu17,S17} have of explaining supernov\ae.  If we didn't know that
core collapse supernov\ae\ actually exist, we would probably conclude that
core collapses lead, depending on the mass of the collapsing star, either to
a black hole or to a darkly formed neutron star.  Decades of work on
hydrodynamics and neutrino transport, together with inclusion of angular
momentum, have led to calculated explosions, but this work has been
motivated, and perhaps implicitly biased, towards that result.  Hence it is
plausible that neutron stars may be formed, without an explosion, from the
core collapses of some stars below the neutron star upper mass limit; 
observationally, the star simply ``winks out''.

The second argument is empirical.  Most Galactic radio pulsars have space
velocities of several hundred km/s \citep{LL94}.  Their spatial distribution
indicates that they are Population I objects, so they must acquire these
velocities when the neutron stars are formed.  Yet there are also pulsars
and neutron star X-ray binaries (in fact, a superabundance) in globular
clusters that have escape velocities of 10--20 km/s \cite{K75}.  Only
$\lesssim 10^{-4}$ of the phase space of the Galactic pulsar velocity
distribution is at speeds low enough for a pulsar, or a binary containing
it, to be retained in the globular cluster.   Additional empirical evidence
for neutron star formation with low recoil has been derived from studies of
the double pulsar PSR J0737-3039 \citep{PS05,DPS14,BP16,BHP16}.

These globular cluster neutron stars must be produced in very different
events, collapses that give very little recoil to the forming neutron star.
An obvious candidate for such events is a dark core collapse; with no mass
ejected, there is no mechanical recoil.  Other recoil-inducing processes
(anisotropic neutrino emission, interference of magnetic dipole and
quadrupole radiation) must also be weak.   It is unclear what properties
of the progenitor determine which path (dark or explosive) its collapse will
take.
\section{Na\"{\i}ve model, caveats and statistics}
The physics of supernova remnants is complex, even in their early phases,
and impossible to predict quantitatively without much better knowledge of
their parameters than is forseeable.  Significant processes include
recombination, shock reionization following collision with surrounding gas
and photoionization by radiation from the neutron star \citep{MBM17,PBS17}.
In order to estimate the feasibility of testing the hypothesis of dark
neutron star formation by comparing it to expectations if a supernova
remnant is present, we adopt a na\"{\i}ve model of the evolution of its
contribution to the dispersion measure.  If the model predicts an observable
variation, then we may consider an absence of such variations in a large
number of repeating FRB as evidence for dark neutron star formation.
If there are also repeating FRB whose dispersion measures do vary
\citep{PBS17}, that would be evidence for a bimodal character of their
formation processes, analogous to the bimodal distribution of neutron star
recoil velocities.  From only one confirmed repeating FRB, it is not
possible to form any firm conclusions, but when scores of repeating FRB are
discovered, it may be possible to form significant conclusions.
\subsection{SNR model}
A na\"{\i}ve model of dispersion by a supernova remnant assumes spherical
symmetry\footnote{Observed supernova remnants are highly asymmetric, both
on large angular scales (spherical harmonic indices 1, 2, 3, {\it etc.\/})
and on fine scales, visible as filaments (spherical harmonic indices
$\gtrsim 100$).  The dependence of electron column density on direction is
unknown, and the na\"{i}ve model describes its average.  The problem is even
more complex if velocities are nonradial, because then a filament can cross
the line of sight as the remnant expands.  Nonradial velocities that,
following a point explosion, can only be produced by asymmetric deposition of
energy or interaction with asymmetrically distributed circumstellar matter,
increase the rate of change of dispersion measure and therefore strengthen
any conclusions inferred from its upper bounds.}.  At early times
interaction with interstellar matter is insignificant, and for a fully
ionized ejecta mass $M \equiv M_{10} \times 10 M_\odot$ of cosmic
composition ($\langle Z/A \rangle = 0.85$; such a massive envelope is likely
hydrogenic) expelled at a speed $v \equiv v_9 \times 10^9\,$cm/s, after a
time $t \equiv t_9 \times 10^9\,$s, the dispersion measure
\begin{equation}
\text{DM} = {3 f_{geom} \over 4 \pi} {M \over m_p v^2 t^2}  \left\langle{Z
\over A} \right\rangle = 787 {f_{geom} M_{10} \over v_9^2 t_9^2}\
\text{pc-cm}^{-3},
\end{equation}
where $f_{geom} = 1$ for a homogeneous sphere (with $v$ the expansion speed
at its surface) and $f_{geom} = 1/3$ for a thin shell.

The remnant contribution to the FRB dispersion measure is not known because
the intergalactic medium makes a substantial, likely dominant, contribution
whose magnitude is unknown (unless the FRB is associated with a galaxy with
measured redshift).  The unknown near-source environment and uncertain
Galactic interstellar medium also contribute.  However, the supernova
remnant is likely the only significant source of the time derivative of the
dispersion measure:
\begin{equation}
\label{DMdot}
\dot{\text{DM}} = -{50 f_{geom} M_{10} \over v_9^2 t_9^3} {\text{pc} \over
\text{cm}^3\,\text{y}},
\end{equation}
or
\begin{equation}
\label{age}
t_9 = \left({50 f_{geom} M_{10} \over \left\vert \dot{\text{DM}} \right\vert
v_9^2}\right)^{1/3}.
\end{equation}
In Eq.~\ref{age} $\left\vert \dot{\text{DM}} \right\vert$ is expressed in
pc/cm$^3$y.  For FRB 121102 $\left\vert \dot{\text{DM}} \right\vert
\lessapprox 2\,$pc/(cm$^3$y) \citep{C17} and
\begin{equation}
\label{agebound}
t_9 \gtrapprox 3 \left({f_{geom} M_{10} \over v_9}\right)^{1/3};
\end{equation}
its age is $\gtrapprox 100\,$y if the dimensionless variables are $\sim 1$.

If FRB 121102 is this old then its rotation rate will not slow by more than
$\sim 30\%$ over the next century, and its FRB activity \emph{may} remain
roughly the same as it is today.  However, we cannot exclude activity
varying (but not systematically decaying) on shorter time scales, as
observed for SGR that are several thousand years old.

Some of the plausible complications \citep{MBM17} of this na\"{\i}ve model
would have the effect of decreasing the bound in Eqs.~\ref{agebound} by
providing additional contributions to $\dot{\text{DM}}$.  For example,
recombination of an ionized remnant would make $\dot{\text{DM}}$ more
negative, mimicking and adding to the effect of expansion, while shock or
photo-ionization would oppose the effect of expansion, possibly leading to
$\dot{\text{DM}} > 0$.  A recombined remnant would contribute nothing to DM
($M_{10} = 0$), and its $\dot{\text{DM}} = 0$.
\subsection{Spindown regimes}
\label{spindown}
An elementary calculation shows that there are two spindown regimes,
depending on the magnitude of the magnetic dipole moment $\mu$.  We assume
that $\mu$ is constant and use the dipole radiation expression with ${\vec
\mu} \perp {\vec \omega}$.  If
\begin{equation}
\label{mu1}
\mu > \sqrt{{3 \over 4}{I c^3 \over \omega^2 t}},
\end{equation}
where $I \approx 10^{45}\,$g-cm$^2$ is the neutron star moment of inertia
and $\omega = \omega_4 \times 10^4/\text{s} < \omega_{birth}$ is its present
angular spin rate.  Integrating the spindown equation yields, if $\omega \ll
\omega_{birth}$, the present spindown power 
\begin{equation}
P = {3 \over 8}{I^2 c^3 \over \mu^2 t^2} < {1 \times 10^{37} \over B_{15}^2
t_9^2}\ \text{ergs/s},
\end{equation}
where $\mu = 10^{33} B_{15}\,$gauss-cm$^3$.  Substituting from
Eq.~\ref{mu1},
\begin{equation}
\label{Pmax1}
P < {1 \over 2}{I \omega^2 \over t} = 2 \times 10^{43} {\omega_4^2 \over
t_\text{100 y}}\ \text{ergs/s}.
\end{equation}

For smaller values of the magnetic moment
\begin{equation}
\label{mu2}
\mu < \sqrt{{3 \over 4}{I c^3 \over \omega^2 t}},
\end{equation}
the spin rate $\omega \approx \omega_{birth}$ and, using Eq.~\ref{mu2},
\begin{equation}
\label{Pmax2}
P = {2 \over 3} {\mu^2 \omega^4 \over c^3} < {1 \over 2}{I \omega^2 \over t}
= 2 \times 10^{43} {\omega_4^2 \over t_\text{100 y}}\ \text{ergs/s},
\end{equation}
identical to Eq.~\ref{Pmax1}.

The unsurprising result that $P < I \omega^2/(2 t)$, where $t$ is the
neutron star's age, is consistent with the power observed for FRB 121102, if
both $\mu$ and $\omega$ be close to their optimal values.  That might seem
implausible, but there is a strong selection effect favoring detection of
sources with optimal parameters\footnote{This is the same argument that
supports the assumption of energy equipartition between field and particles
in incoherent synchrotron sources.}.  These conditions are relaxed if the
radiation is beamed \citep{K17a} or if energy storage \citep{K17b} frees FRB
from the limit of the spindown power. The ages of FRB that have not been
observed to repeat is unknown, and may be very much shorter, permitting much
greater powers.
\section{Statistics of $\dot{\text{DM}}$}
\label{statistics}
When many repeating FRB are discovered statistical inference may become
possible.  From Eq.~\ref{DMdot} we find the distribution of
$\dot{\text{DM}}$:
\begin{equation}
\label{ddotDM}
{d\dot{\text{DM}} \over dt} = 4.7 {f_{geom} M_{10} \over v_9^2 t_9^4}
{\text{pc} \over \text{cm}^3\,\text{y}^2}.
\end{equation}

If the intrinsic rate of FRB activity does not change as the supernova
remnant ages and expands (so there is no selection effect favoring younger
objects) then the distribution of $\dot{\text{DM}}$ is
\begin{equation}
{dN\left(\dot{\text{DM}}\right) \over d\dot{\text{DM}}} \propto {dt \over
d\dot{\text{DM}}} = {39 \over \left(-\dot{\text{DM}}\right)^{4/3}} \left({
f_{geom} M_{10} \over v_9^2}\right)^{1/3} \left({\text{pc\,y}^2 \over 
\text{cm}^3}\right)^{1/3}.
\end{equation}

The assumptions (no evolution of the FRB, the na\"{i}ve SNR model and the
absence of any correlation between FRB properties and the SNR parameters)
may be tested with a measured distribution of $\dot{\text{DM}}$.  Nonzero
$\left\vert \dot{\text{DM}} \right\vert$ is detectable if
$\left\vert \dot{\text{DM}} \right\vert > \left\vert \dot{\text{DM}}
\right\vert_{thresh}$.  The detection threshold $\left\vert \dot{\text{DM}}
\right\vert_{thresh}$ is determined by the accuracy of DM measurements, the
number of bursts observed, and the duration of the observational baseline.
$\left\vert \dot{\text{DM}} \right\vert_{thresh} \approx 2$ pc/cm$^3$y with
about three years of data.

This corresponds to a maximum age $t < 100 (f_{geom} M_{10}/v_9^2)^{1/3}\,$y
(Eq.~\ref{age}) for which $\left\vert \dot{\text{DM}} \right\vert >
\left\vert \dot{\text{DM}} \right\vert_{thresh}$.  If FRB are active for a
lifetime $T < t$, then all should show measurable non-zero $\dot{\text{DM}}$
(in contrast to FRB 121102).  If $T > t$ the fraction
\begin{equation}
F(\dot{\text{DM}} < 0) = {t \over T} \approx {100\,\text{y} \over T}
\left({f_{geom} M_{10} \over v_9^2 \left\vert \dot{\text{DM}}
\right\vert_{thresh}}\right)^{1/3}
\end{equation}
are predicted to show measurable non-zero $\left\vert \dot{\text{DM}}
\right\vert > \left \vert \dot{\text{DM}} \right\vert_{thresh}$.  This is
a testable consequence of the na\"{\i}ve supernova remnant model, and its
empirical disproof would be evidence of dark neutron star formation.

Accidental cancellation of $\dot{\text{DM}} < 0$ by a positive contribution
(ionization of a recombined remnant) is possible, yielding $\left\vert
\dot{\text{DM}} \right\vert < \left\vert \dot {\text{DM}}
\right\vert_{thresh}$, but unlikely.  If it occurs then some other FRB would
be expected to have $\dot{\text{DM}} > 0$ because accurate cancellation of
two contributions of opposite signs must be fortuitous.

If FRB are isotropic radiators without energy storage, then $T$ is limited
\citep{K17a} by the requirement of producing peak powers that for some FRB
are $\sim 10^{43}\,$ergs/s.  Using the most optimistic estimates from
Sec.~\ref{spindown}, $T \lessapprox 200\,$y, and this argument predicts that
$\gtrapprox 50\%$ of repeating FRB should show nonzero $\dot{\text{DM}}$ in
a few years of observation.  If this is not so, which would be established
at 95\% significance if the first five repeaters with a few years of data
all showed $\dot{\text{DM}} = 0$ to the accuracy of measurement, then one or
more of the assumptions of the na\"{\i}ve model would be falsified.  We
suggest as a candidate to be rejected the assumption $M_{10} \ne 0$; if
$M_{10} = 0$ then $\dot{\text{DM}} = 0$.  This would imply either dark
neutron star formation, with no supernova remnant (or supernova), or a
recombined remnant.
\section{LSST---Very Recent SN?}
The Large Synoptic Survey Telescope \citep{LSST} will accumulate a decade
of observations over more than half the sky with a cadence of about once
per two days (season and moonlight permitting) to a magnitude of $r \sim
27.5$.  This will be sufficient to detect Type II supernov\ae\ (of absolute
magnitude -17 and apparent magnitude about 25.5 at a luminosity distance of
3 Gpc) if they are at the arc-second position of a repeating FRB, determined
interferometrically.  Searching for such supernov\ae\ will test the
hypothesis of dark FRB formation within the epoch of the LSST database.

The discovery of such a supernova would immediately determine the age of
the neutron star making the FRB, and would disprove (at least for that FRB)
the hypothesis of dark formation.  That hypothesis is falsifiable, but its
truth cannot be demonstrated.
\section{Discussion}
The constancy over more than three years of the dispersion measure of the
repeating FRB 121102 places strong constraints on any supernova remnant
surrounding it.  In a na\"{\i}ve model of a core collapse supernova remnant
this leads to an approximate lower bound on its age of $\sim 100$ y.  A
similarly na\"{i}ve model of FRB as giant pulsar pulses, implausibly
assuming 100\% efficient conversion of rotational energy to the coherent FRB
pulse, indicates ages of the neutron star in more energetic FRB of $\lesssim
200\,$y and $\lesssim 2 \times 10^4\,$y in FRB 121102.  More plausible
values $\ll 1$ of efficiency lead to contradictions with ages implied by the
na\"{i}ve supernova remnant model.

These numbers are sufficiently uncertain that contradiction is avoidable,
provided that implausibly high (much higher than observed in pulsars)
efficiency of conversion of rotational energy to coherent radiation
is allowed, or other loopholes (narrowly beamed radiation, energy storage)
are considered.  Still, the constancy of the dispersion measure of FRB
121102 suggests that no supernova remnant surrounds it.  That would imply
dark neutron star formation, a hypothesis that makes other testable
predictions: the constancy of the dispersion measure of other repeating FRB
and the absence of supernov\ae\ in LSST observations of the locations of
FRB.

\bsp 
\label{lastpage} 
\end{document}